\documentclass{emulateapj}

\usepackage{graphicx}

\newcommand{\eg}{e.g., }
\newcommand{\ie}{i.e., }

\newcommand{\kms}{km~s$^{-1}$}

\def\gsim{\mathrel{\rlap{\lower 4pt \hbox{\hskip 1pt $\sim$}}\raise 1pt
\hbox {$>$}}}
\def\lsim{\mathrel{\rlap{\lower 4pt \hbox{\hskip 1pt $\sim$}}\raise 1pt
\hbox {$<$}}}
\newcommand{\vph}{v_{\rm ph}}
\newcommand{\td}{t_{\rm d}}

\newcommand{\vcl}{v_{\rm cl}}
\newcommand{\acl}{\alpha_{\rm cl}}
\newcommand{\fcl}{f_{\rm cl}}

\newcommand{\pasa}{PASA}

\def\ion#1#2{{\rm #1}~{\sc #2}}

\shorttitle{3D Explosion Geometry of Stripped-Envelope Core-Collapse SNe II}
\shortauthors{Tanaka et al.}

\begin{document}

\title{
Three-Dimensional Explosion Geometry of Stripped-Envelope Core-Collapse Supernovae. II. Modelling of Polarization}
\author{
  Masaomi Tanaka\altaffilmark{1},
  Keiichi Maeda\altaffilmark{2,3},
  Paolo A. Mazzali\altaffilmark{4},
  Koji S. Kawabata\altaffilmark{5,6},
  Ken'ichi Nomoto\altaffilmark{3,7}
}

\altaffiltext{1}{National Astronomical Observatory of Japan, Osawa, Mitaka, Tokyo 181-8588, Japan; masaomi.tanaka@nao.ac.jp}
\altaffiltext{2}{Department of Astronomy, Kyoto University, Kitashirakawa-Oiwake-cho, Sakyo-ku, Kyoto 606-8502, Japan}
\altaffiltext{3}{Kavli Institute for the Physics and Mathematics of the Universe (WPI), The University of Tokyo, Kashiwanoha, Kashiwa, Chiba 277-8583, Japan}
\altaffiltext{4}{Astrophysics Research Institute, Liverpool John Moores University, Liverpool L3 5RF, UK}
\altaffiltext{5}{Hiroshima Astrophysical Science Center, Hiroshima University, Kagamiyama, Higashi-Hiroshima, Hiroshima 739-8526, Japan}
\altaffiltext{6}{Department of Physical Science, Hiroshima University, Kagamiyama, Higashi-Hiroshima 739-8526, Japan}
\altaffiltext{7}{Hamamatsu Professor}

\begin{abstract}
We present modelling of line polarization to study 
multi-dimensional geometry of stripped-envelope core-collapse supernovae (SNe).
We demonstrate that a purely axisymmetric, two-dimensional geometry 
cannot reproduce a loop in the Stokes $Q-U$ diagram, \ie
a variation of the polarization angles along the velocities
associated with the absorption lines.
On the contrary, three-dimensional (3D) clumpy structures naturally 
reproduce the loop.
The fact that the loop is commonly observed in stripped-envelope SNe 
suggests that SN ejecta generally have a 3D structure.
We study the degree of line polarization as a function of the
absorption depth for various 3D clumpy models with different
clump sizes and covering factors.
Comparison between the calculated and observed degree of line polarization
indicates that a typical size of the clump 
is relatively large, $\gsim 25 \%$ of the photospheric radius.
Such large-scale clumps are similar to those observed in 
the SN remnant Cassiopeia A.
Given the small size of the observed sample, 
the covering factor of the clumps is only weakly constrained ($\sim 5-80 \%$).
The presence of large-scale clumpy structure suggests 
that the large-scale convection or standing accretion shock instability
takes place at the onset of the explosion.
\end{abstract}


\section{Introduction}
\label{sec:intro}

\begin{figure*}[t]
\begin{center}
\begin{tabular}{ccc}
\includegraphics[scale=0.85]{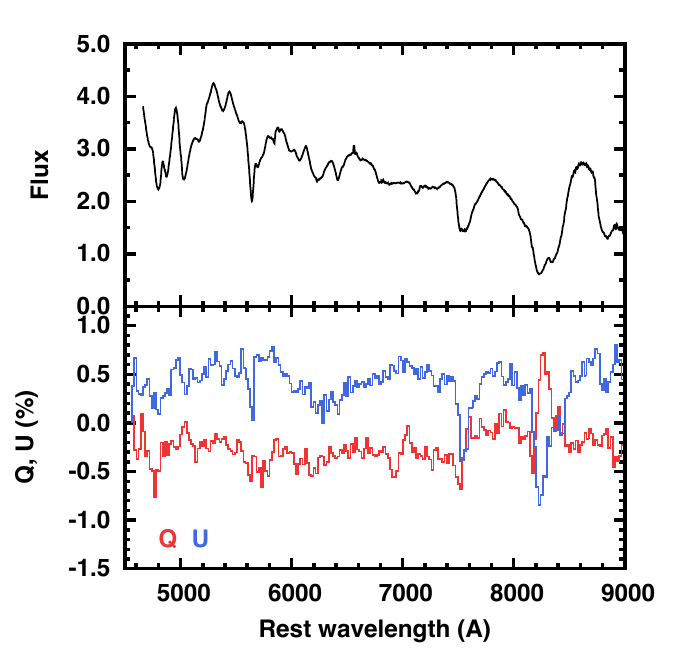} &
\includegraphics[scale=0.85]{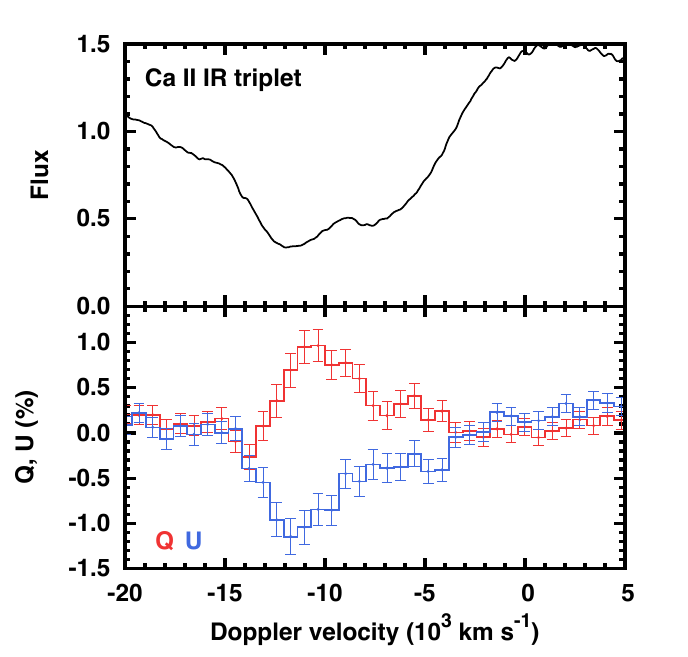} &
\includegraphics[scale=0.85]{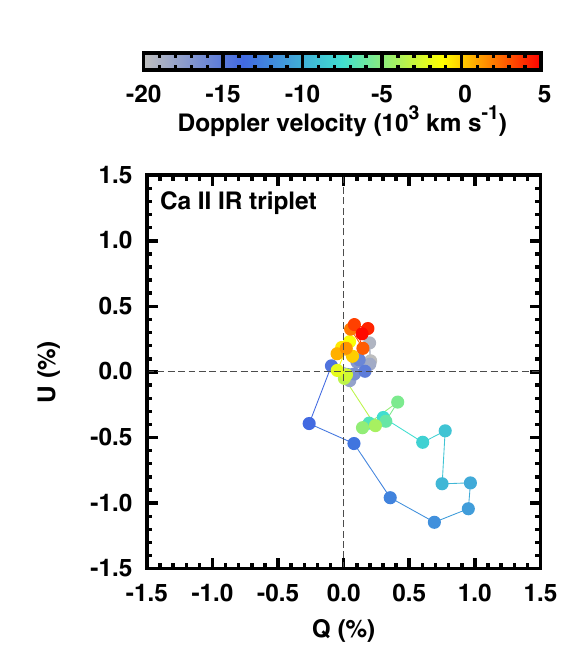} \\
\end{tabular}
\caption{
Example of observed spectropolarimetric data
(Type Ib SN 2009jf, \citealt{tanaka12pol}).
{\it Left}: Flux spectrum (top) and polarization spectrum (bottom).
{\it Middle}: The same data around the \ion{Ca}{ii} IR triplet 
line as a function of Doppler velocity. 
{\it Right}: The polarization data around the \ion{Ca}{ii} IR triplet line 
in the $Q-U$ diagram.
In the middle and right panels,
an estimated interstellar polarization 
($Q=-0.25 \%$ and $U=0.30 \%$, constant over the narrow wavelength range
around the \ion{Ca}{ii} line) has been subtracted.
\label{fig:09jf}}
\end{center}
\end{figure*}

Core-collapse supernovae (SNe) are the explosions of massive stars.
Since core-collapse SNe eject heavy elements
synthesized inside of the stars,
they play vital roles in chemical enrichment of galaxies.
In addition, because of the large kinetic energy of
the explosion ($\sim 10^{51}$ erg),
SNe are also important for the galaxy formation.
Despite their importance,
the mechanism of the core-collapse SNe
is a long-standing mystery 
(see \citealt{janka12,kotake12,burrows13,mueller16} for reviews).
Results of numerical simulations agree
to the point that massive stars would not explode in one dimensional simulations
\citep{rampp00,liebendorfer01,thompson03,sumiyoshi05}
except for some cases of the least massive stars \citep{kitaura06,janka08}.
Therefore, multi-dimensional effects or deviation from spherical
symmetry are believed to be crucial for successful explosions. 

The leading scenario of core-collapse SNe is
neutrino-driven explosion, where
multi-dimensional effects can appear by convection 
\citep[\eg][]{herant94,burrows95,janka96}
or standing accretion shock instability 
\citep[SASI, \eg][]{blondin03,scheck04,ohnishi06,foglizzo07,ott08,iwakami08,fernandez10,hanke12}.
In fact, some successful explosions have been reported
by two-dimensional (2D) or three-dimensional (3D) simulations
\citep[\eg][]{buras06,marek09,suwa10,mueller12,takiwaki12,hanke13,bruenn13,takiwaki14,couch14,melson15,lentz15,roberts16,mueller16}
although the obtained explosion energy is usually lower than $10^{51}$ erg.
Another scenario is magneto-rotational explosion 
\citep[\eg][]{yamada04,kotake04,takiwaki09,sawai05,burrows07,obergaulinger09,moesta14},
where the amplified magnetic fields drive the explosion.
In this scenario, the bipolar explosion is generally expected.

In order to link these theoretical models with observations,
it is necessary to derive
multi-dimensional geometry from observed SNe.
The most straightforward method is spatially-resolved
observations of nearby SN remnants 
\citep[\eg][]{hwang04,isensee10,delaney10,milisavljevic15}.
However, the number of accessible objects is limited.
To advance our knowledge,
it is therefore important to study the multi-dimensional 
geometry of extragalactic SNe.
In fact, many efforts have been made to derive
the multi-dimensional geometry from extragalactic SNe,
for example by using spectral line profiles at late phases
($\gsim$ 1 yr after the explosion, 
\eg \citealt{spyromilio94,sollerman98,matheson00,mazzali01,maeda02,elmhamdi04,mazzali05,maeda08,modjaz08,tanaka0908Dneb,taubenberger09,maurer10,chornock10IIP,valenti11,shivvers13,roy13,chen14,milisavljevic1512ap,mauerhan16})

Polarization at early phases ($\lsim$ 50 days after the explosion)
is one of the most powerful methods to derive multi-dimensional
geometry from extragalactic SNe (see \citealt{wang08} for a review).
By observations, we can measure continuum and line polarizations.
In the SN ejecta, electron scattering is the dominant source of polarization.
Line scattering generally produces less polarization
(\citealt{howell01,kasen03}, see also \citealt{jeffery89}),
and it is often assumed that line scattering works as a de-polarizer.
From the spherical symmetric SN ejecta,
no polarization should be detected because of complete cancellation
of polarization vectors.
Non-zero continuum polarization would be observed
when the photosphere deviates from spherical symmetry
\citep{shapiro82,hoeflich91,hoeflich96,dessart11pol,bulla16}.
In addition, even for the spherical photosphere,
non-zero line polarization would be observed 
when the distribution of an ion producing the corresponding absorption 
line is not spherical symmetry \citep{kasen03,hole10}.
Therefore, line polarization can be a diagnostic to multi-dimensional
element distribution in the SN ejecta.

In this paper, we present modelling of line polarization
in stripped-envelope SNe (SNe of Type IIb, Ib, and Ic)
to obtain connections between the polarization properties
and the element distribution in the SN ejecta.
Compared with the cases of H-rich SNe, 
closer insight on the explosion mechanism can be obtained
for stripped-envelope SNe,
as the large hydrogen envelope is not present.
In Section \ref{sec:methods}, we describe our method to compute
polarization signature of the SN models.
In Sections \ref{sec:2D} and \ref{sec:3D},
we show results of 2D and 3D models, respectively.
We discuss implication of our results in Section \ref{sec:discussion}
and give summary in Section \ref{sec:summary}.

\section{Methods}
\label{sec:methods}

\subsection{Radiation Transfer}

We perform 3D radiation transfer simulations
to study the properties of the line polarization.
For this purpose, we use a simple
3D Monte Carlo radiation transfer code.
The code takes into account the electron
scattering and the line scattering.
We treat only a {\it single} line at a {\it single} epoch
rather than modelling time evolution of full spectra
since we aim to obtain the connection
between explosion geometry and properties of line polarization
(see \citealt{hole10} for a similar strategy).
The code computes the polarization spectrum of the line
for arbitrary 3D distribution of the line optical depth.
More details of the code are given in Appendix.

We use $100 \times 100 \times 100$ linearly distributed Cartesian meshes.
The velocity is used as a spatial coordinate 
thanks to the homologous expansion ($r \propto v$).
The maximum velocity is 25000 \kms, and thus
the resolution is 500 \kms, giving the resolution
of $\lambda/ \Delta \lambda = c/ \Delta v = 600$, 
which is comparable to a typical spectral resolution of
low-resolution spectropolarimetric observations.

We start simulations by generating
unpolarized photon packets from the spherical inner boundary
($v = v_{\rm in}$).
The electron scattering optical depth from the inner boundary
to infinity is set to $\tau_{\rm in}$.
In this paper, we adopt $\tau_{\rm in} = 3$ 
as in \citet{kasen03} and \citet{hole10}.
Note that the photosphere ($v = v_{\rm ph}$) is defined as the position
where the electron scattering optical depth is unity, and thus,
the inner boundary of the computation is located inside of the photosphere.

The photon packets are then tracked by taking into 
account the electron scattering and the line scattering.
For the electron scattering, we use a power-law 
electron density profile, $n_e \propto r^{-n}$.
The electron density is assumed to be spherically symmetric.
For the power-law index, we use $n = 7$
which describes the line forming region of
hydrodynamic models of stripped-envelope SNe \citep{iwamoto00,mazzali0097ef}.
Although the very outermost ejecta has a steeper slope
($n \sim 10$, \citealt{matzner99}),
we use a single power-law profile
since the outermost ejecta does not have a strong contribution
to absorption lines.
Note that polarization pattern is not affected by the slope
if the slope is steep enough ($n \gsim 5$, \citealt{kasen03}).
We assume the photospheric velocity ($\vph$) and 
the time after the explosion ($t$),
which give the photospheric radius $r_{\rm ph} = \vph t$.
Then, with the condition that the electron scattering optical depth
is unity at the photosphere,
the normalization of the electron density is obtained.
We adopt $\vph = 8,000$ \kms\ and $t = 20$ days as typical values
for stripped-envelope SNe around the maximum light.

For the line scattering, we use Sobolev approximation \citep{castor70},
which is a sound approximation in the SN ejecta
with a large velocity gradient.
For the Sobolev line optical depth,
we assume a power-law radial profile above the photosphere,
$\tau_{\rm line} = \tau_{\rm ph} ( r / r_{\rm ph} )^{-n}$.
Here $\tau_{\rm ph}$ is the Sobolev optical depth at the photosphere.
For simplicity, we use the same power-law index with 
the electron density ($n = 7$).

In addition to the spherical component of the line optical depth,
we assume an enhancement by a factor of $f_{\rm \tau}$ in some regions
\eg a torus or clumps.
Note that this is different from the treatment by \citet{hole10},
where the line opacity is set to be zero outside of the clumps.
Such a treatment seems more suitable for Type Ia SNe
(as they applied for),
where a strong line is formed dominantly in a certain layer,
\eg Si lines are produced mostly in the Si-rich layers.
On the other hand, for strong lines in stripped-envelope
core-collapse SNe, such as those of Ca and Fe,
both pre-SN and newly synthesized elements contribute to the absorption.
Therefore, we assume an enhancement in addition to the spherical component.
In the models presented in this paper, we adopt $f_{\rm \tau} = 10.0$.
Implication of this choice is discussed in Section \ref{sec:3D}.
The parameters for the models are summarized
in Table \ref{tab:param}.

\subsection{Comparison with Observations}

We study the explosion geometry of stripped-envelope SNe by comparing 
results of our simulations with observations.
Figure \ref{fig:09jf} shows an example of 
spectropolarimetric data of stripped-envelope SNe 
(Type Ib SN 2009jf, \citealt{tanaka12pol}).
In this paper, we define Stokes parameters as a fraction of the total flux:
$Q \equiv \hat{Q}/I$ and $U \equiv \hat{U}/I$, where
$\hat{Q}$ and $\hat{U}$ are polarized fluxes,
\ie $\hat{Q} = I_{0} - I_{90}$ and
$\hat{U} = I_{45} - I_{135}$, respectively
($I_{\psi}$ is the intensity measured through the
ideal polarization filter with an angle $\psi$).
From Stokes parameters $Q$ and $U$,
the position angle of the polarization, $\theta$, is obtained by 
$2\theta = {\rm atan}(U/Q)$.

Properties of line polarization in stripped-envelope SNe
can be summarized as follows:
\begin{enumerate}
\item
Non-zero line polarization is common 
and polarization feature shows an 
inverted P-Cygni profile which peaks at flux absorption minimum
\citep[\eg][]{kawabata02,leonard0202ap,wang0302ap,maund0705bf,maund0701ig,maund09,tanaka0807gr,tanaka0905bf,tanaka12pol,mauerhan15,stevance16,mauerhan16}.
\item
When the polarization data across the line 
(middle panel of Figure \ref{fig:09jf}) are plotted
in the Stokes $Q$-$U$ diagram (right panel),
the observed data commonly show a loop 
\citep[\eg][]{maund0705bf,maund0701ig,maund09,tanaka12pol,mauerhan15,stevance16,mauerhan16}.
\footnote{Such a loop in the $Q-U$ diagram has also been observed 
in Type Ia SNe \citep[\eg][]{wang0301el,kasen03,chornock08,patat09,tanaka10,porter16,milne16}
and Type IIn SNe \citep[\eg][]{hoffman08} as well as in Wolf-Rayet stars 
\citep[\eg][]{schulte-ladbeck90,st-louis12}.}
\item
The degree of line polarization, \ie the maximum polarization level at the absorption lines,
is generally a few percent,
and tends to be higher for stronger lines \citep{tanaka12pol}.
\end{enumerate}

The degree of polarization depends on the strength of the absorption.
The absorption strength is mainly determined by global properties of SNe, 
such as ejecta mass, temperature and element abundances,
and not directly by the explosion geometry.
Therefore, it is important to compare features with
a similar absorption strength to discuss the explosion geometry.
Here we define a fractional depth (FD)
of absorption at the absorption minimum,
FD $= (f_{\rm cont} - f_{\rm abs})/f_{\rm cont}$,
where $f_{\rm abs}$ and $f_{\rm cont}$ are
the flux at the absorption minimum and
at the continuum near the absorption line, respectively.
\citet{tanaka12pol} showed that, in a simple configuration, 
the observed polarization ($P_{\rm obs}$) can be approximately described as 
$P_{\rm obs} \simeq P_{\rm cor} [{\rm FD}/({\rm 1-FD})]$,
where a corrected polarization $P_{\rm cor}$ is defined 
as the polarization level if the fractional depth would be FD $=0.5$.

\begin{deluxetable}{lccc} 
\tablewidth{0pt}
\tablecaption{Summary of the models}
\tablehead{
  Model           & $\tau_{\rm ph}$$^{a}$  & $\acl$$^{b}$  & $\fcl$$^{c}$ 
}
\startdata
2D-bipolar-30deg$^d$ &  10.0                &   --        & 0.13 \\
2D-torus-20deg$^e$   &  10.0                &   --        & 0.35 \\ 
\\
3D-a0.5-f0.3    &  3.0,10.0,30.0,100.0  &     0.5    & 0.3 \\
3D-a0.25-f0.3   &  3.0,10.0,30.0,100.0  &     0.25   & 0.3 \\
3D-a0.125-f0.3  &   3.0,10.0,30.0,100.0   &     0.125  & 0.3 \\
\\
3D-a0.5-f0.06  & 30.0                 &    0.5  & 0.06 \\ 
3D-a0.5-f0.2   & 30.0                 &    0.5  & 0.2 \\
3D-a0.5-f0.5   & 10.0                 &    0.5  & 0.4 \\
3D-a0.5-f0.7   & 10.0                 &    0.5  & 0.7 \\
\enddata
\tablecomments{
  $^a$ Sobolev line optical depth at the photosphere.
  $^b$ Size parameter of the clumps for 3D models ($\acl = \vcl/\vph$).
  $^c$ Covering factor of the clumps.
  $^d$ 2D model with the two polar blobs with the half opening angle of 30 deg.
  $^e$ 2D model with an equatorial torus with the half opening angle of 20 deg.
}
\label{tab:param}
\end{deluxetable}


\begin{figure*}[t]
\begin{center}
\begin{tabular}{cc}
  \includegraphics[scale=0.55]{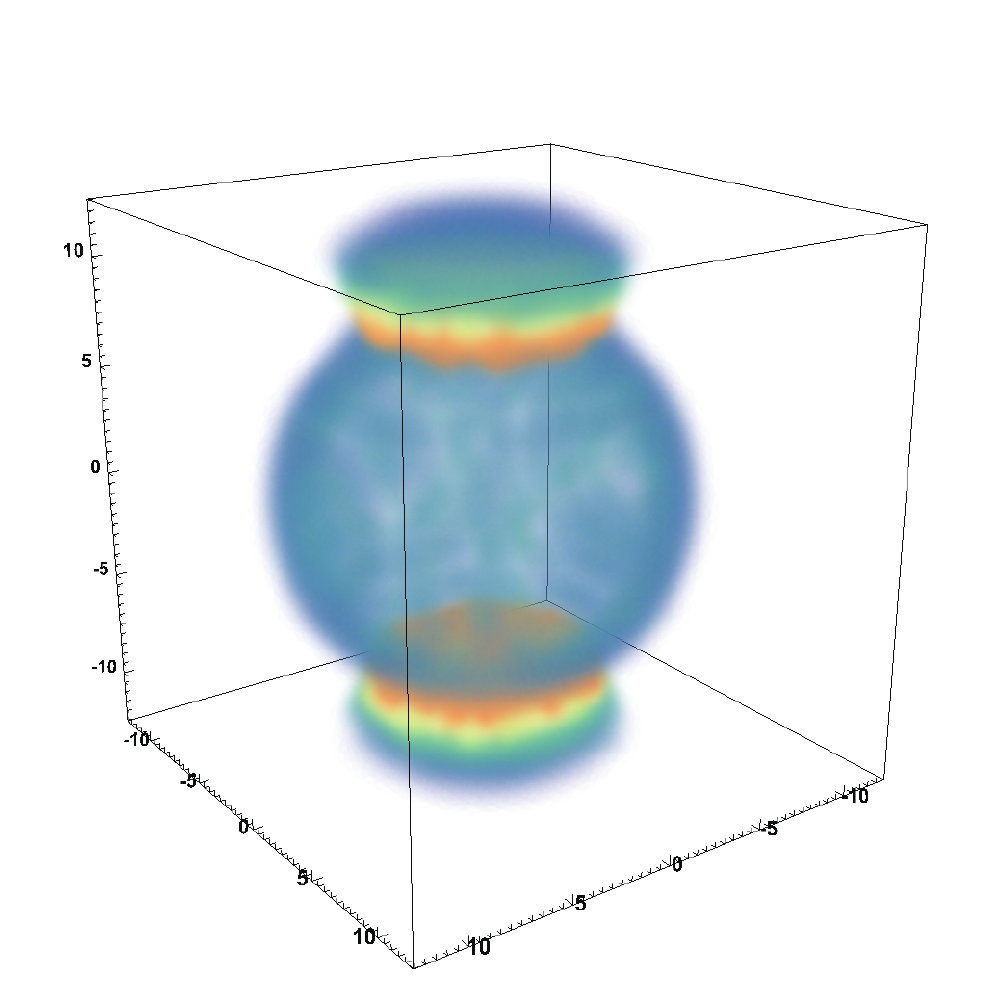} & 
\includegraphics[scale=0.85]{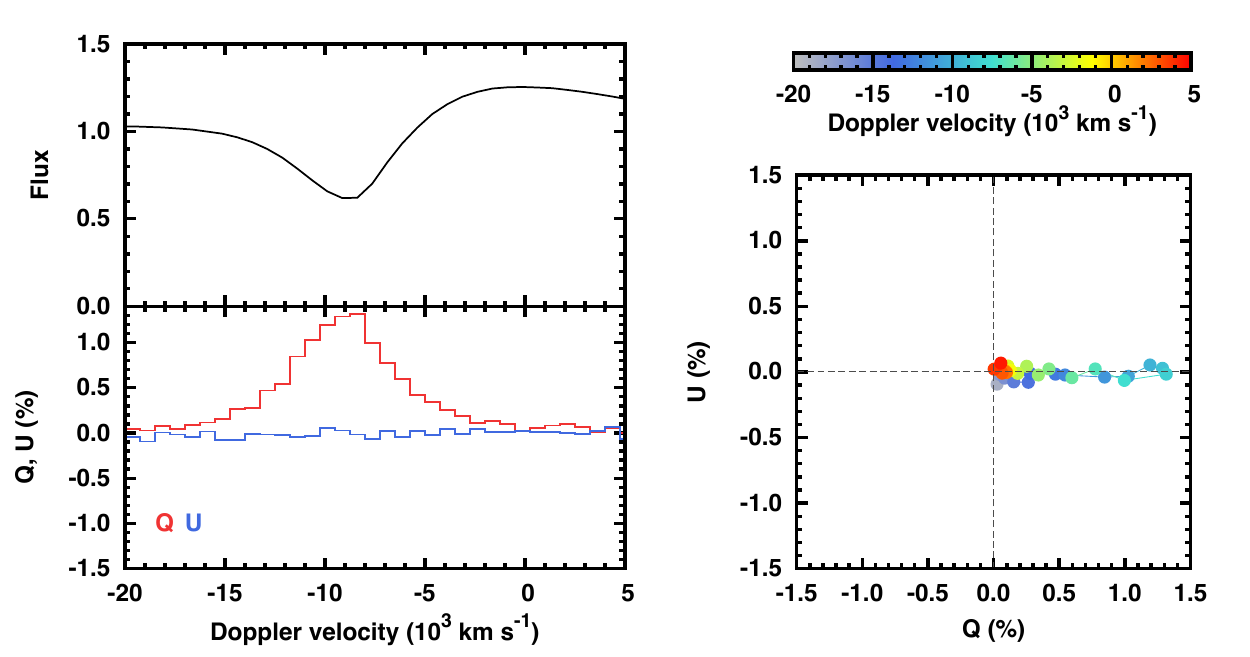}  \\
\includegraphics[scale=0.55]{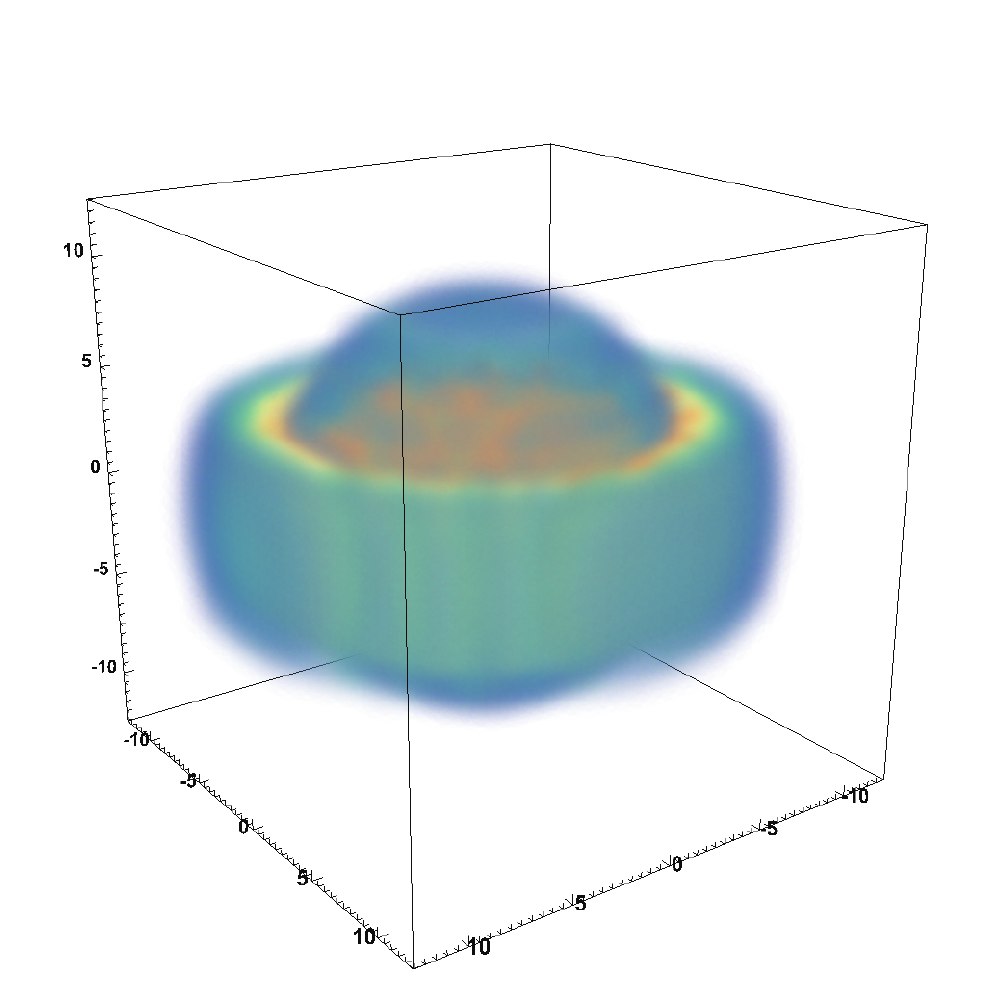} &  
\includegraphics[scale=0.85]{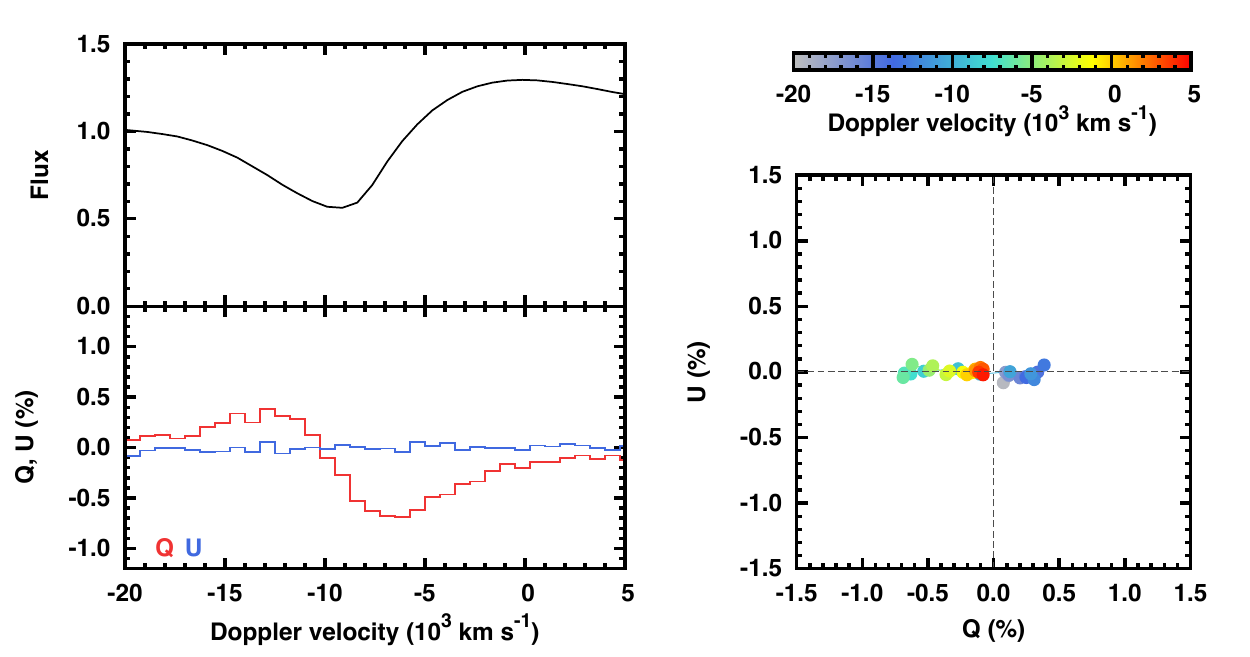}  
\end{tabular}
\caption{
  {\it Top}: Distribution of optical depth for the 2D bipolar model
  (left, 2D-bipolar-30deg),
the simulated polarization spectrum as a function of 
Doppler velocity (middle), and in the $Q-U$ diagram (right).
{\it Bottom}: Same with the top panels but for the 2D torus model
(2D-torus-20deg).
For the optical depth distribution (north is up, east is left),
orange/yellow region shows a higher optical depth
($\tau_{\rm line} \gsim 10.0$)
while green/blue region shows a lower optical depth
($\tau_{\rm line} \lsim 10.0$).
For the simulated polarization spectrum, 
a line of sight is set to be 60 deg from the pole
with the symmetric axis of the model pointing to north.
\label{fig:2D}}
\end{center}
\end{figure*}

\section{Results: 2D Models}
\label{sec:2D}

We first study polarization properties of 
2D axisymmetric models.
As 2D models, we construct a bipolar model and a torus model.
These models are motivated by the results of 
nucleosynthesis calculations for 2D bipolar (or jet-like)
explosion models
\citep[\eg][]{nagataki97,maeda03,nagataki06,tominaga09}.
In these models, explosively synthesized elements such as Fe
are preferably produced in the polar region,
Our 2D bipolar model depicts such a case.
In contrast, the elements produced mainly in the pre-SN stage such as O 
may be distributed in a torus-like geometry,
which are represented by our torus model.

Polarization properties of the bipolar model are shown 
in the top panels of Figure \ref{fig:2D}.
The half opening angle of the polar blobs is set to be 30 deg.
As the opacity distribution is not spherically symmetric, 
non-zero polarization appears.
The polarization data in the $Q-U$ diagram shows a straight line.
This is always the case for every line of sight.
The position in the $Q-U$ diagram represents the position angle,
\ie $\theta = (1/2) {\rm atan}(U/Q)$.
Therefore, the straight line in the $Q-U$ diagram 
means a constant position angle across the P-Cygni profile.

The observed position angle can be rotated depending on the
direction of the symmetric axis of the model on the sky.
However, as long as the 2D bipolar structure is kept, 
the polarization always shows a straight line in 
the $Q-U$ diagram.
Also, this behavior does not depend on global parameters
such as the optical depth at the photosphere ($\tau_{\rm ph}$),
the enhancement factor ($f_{\rm \tau}$)
since this behavior is purely caused by the geometric effect.
We also test the models with different sizes of the blobs
(i.e., opening angles of the bipolar structure),
and confirm that, although the number of lines of sight
to have a high polarization degree depends on the size of the blob,
the straight line in the Q-U plane is always obtained.

Similar polarization properties are obtained for the torus model, 
\ie polarization always shows a straight line in the $Q-U$ diagram.
The bottom panels of Figure \ref{fig:2D} show an example of 
the results for the torus model with a half opening angle of 20 deg.
Note that, for a certain line of sight,
a 90 deg rotation in the position angle can be observed.
For example, for the line of sight of 60 deg from the pole
as shown in Figure \ref{fig:2D},
the lateral part of the photospheric disk is hidden near the photospheric
velocity while the bottom part of the photospheric disk 
is hidden at higher velocities.
As a result, a positive Stokes $Q$ is obtained near the photospheric velocity
while a negative Stokes $Q$ is obtained at higher velocities (middle panel).
This corresponds to a 90 deg rotation in the position angle.
However, only a 90 deg rotation can occur 
as long as the underlying model keeps axisymmetry
since there is no way to produce Stokes $U$ component if the axisymmetric
angle of the model is set to be north ($\theta = 0$ deg) 
as shown in Figure \ref{fig:2D}.
If the symmetric axis of the model is rotated on the sky,
Stokes $U$ components can appear
but the polarization data still form a straight line in the $Q-U$ diagram.

In summary, we validate the statement
commonly made by previous works \citep[\eg][]{kasen03,wang08},
\ie a purely axisymmetric element distribution
cannot reproduce the loop in the $Q-U$ diagram.
When the element distribution has a
purely 2D axisymmetric structure such as bipolar blobs or a torus,
the polarization shows a straight line in the $Q$-$U$ diagram.

\begin{figure*}
\begin{center}
\begin{tabular}{cc}
  \includegraphics[scale=0.55]{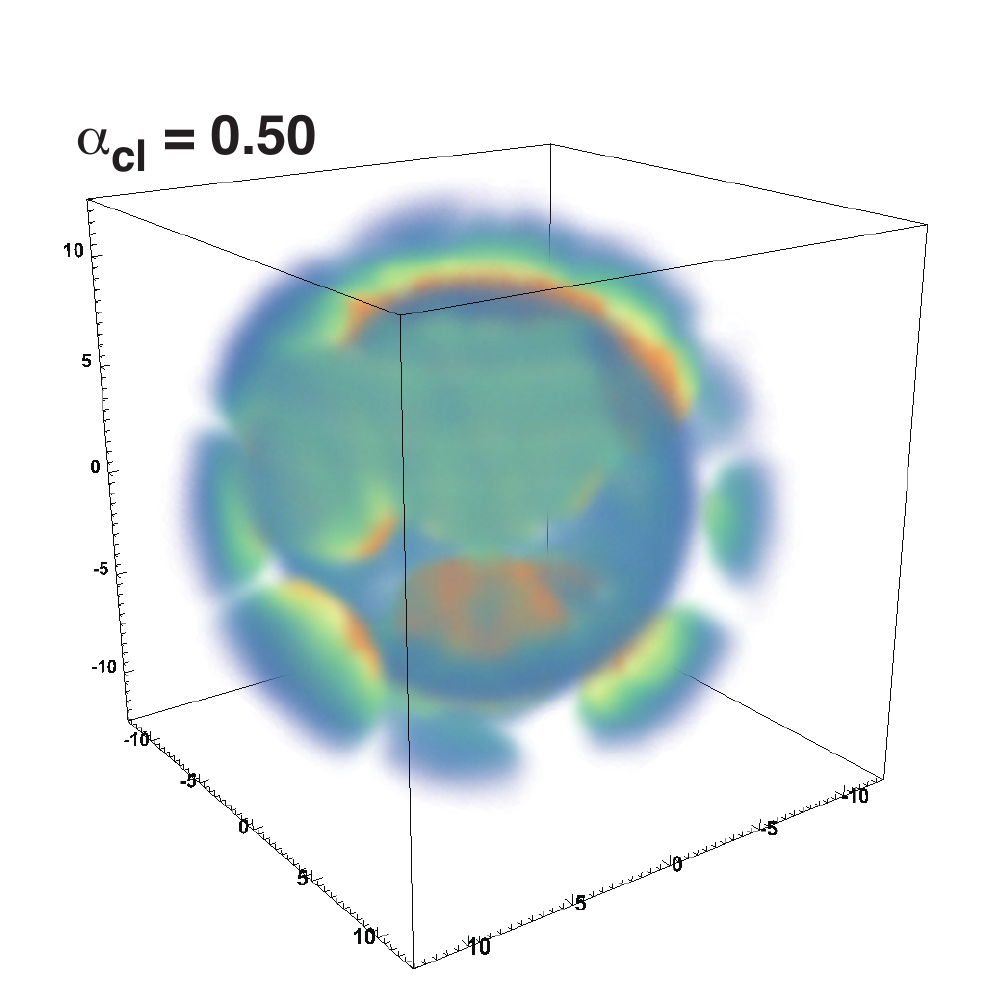} & 
\includegraphics[scale=0.85]{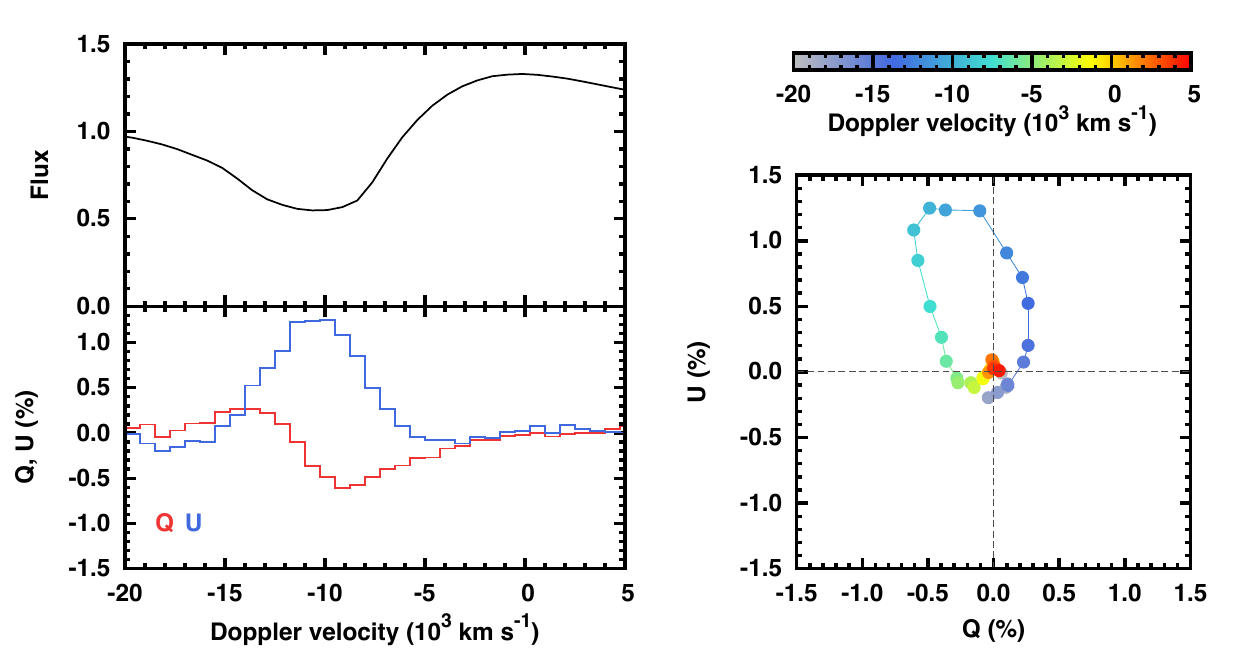}  \\
\includegraphics[scale=0.55]{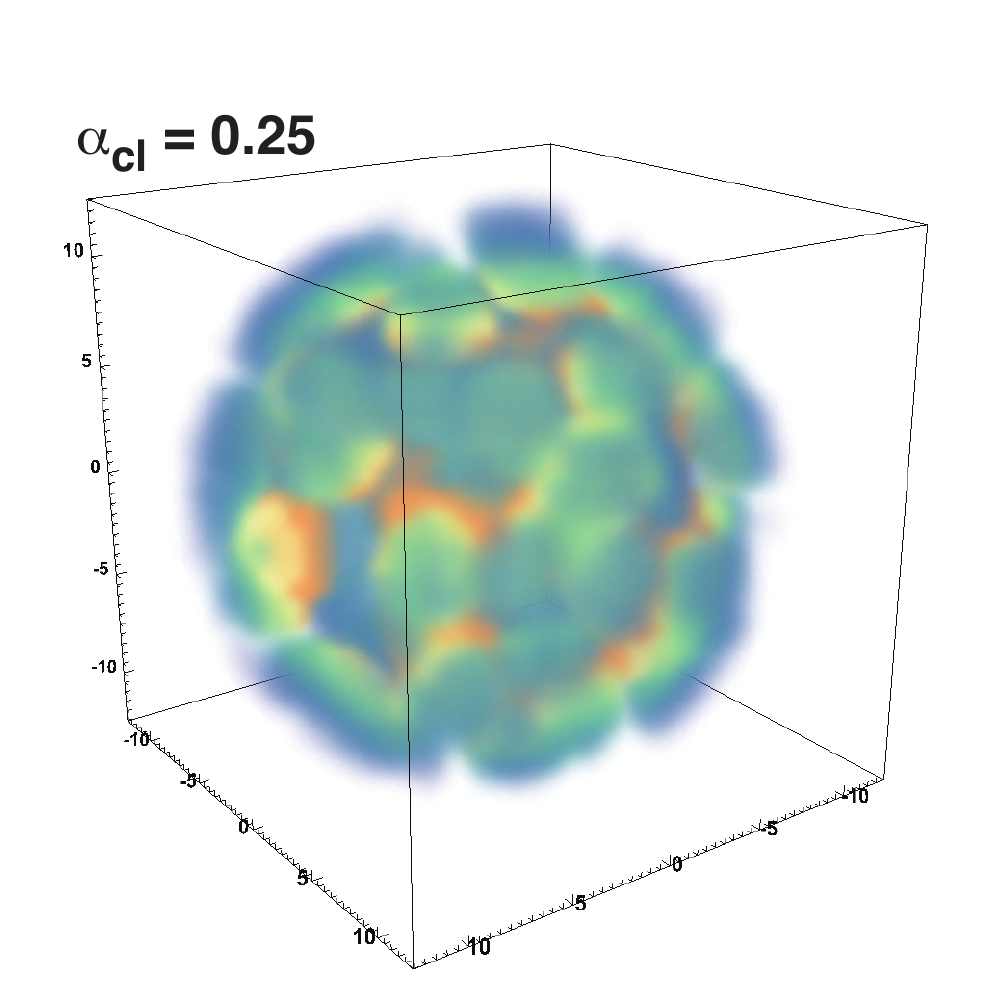} &
\includegraphics[scale=0.85]{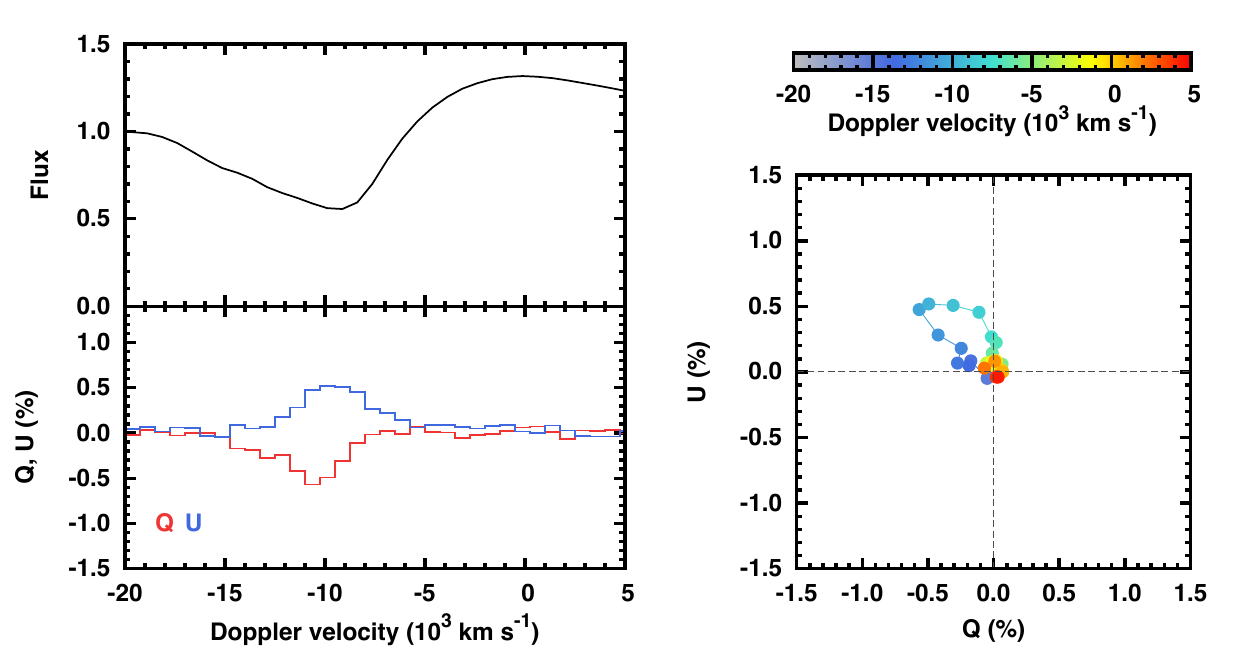}  \\
\includegraphics[scale=0.55]{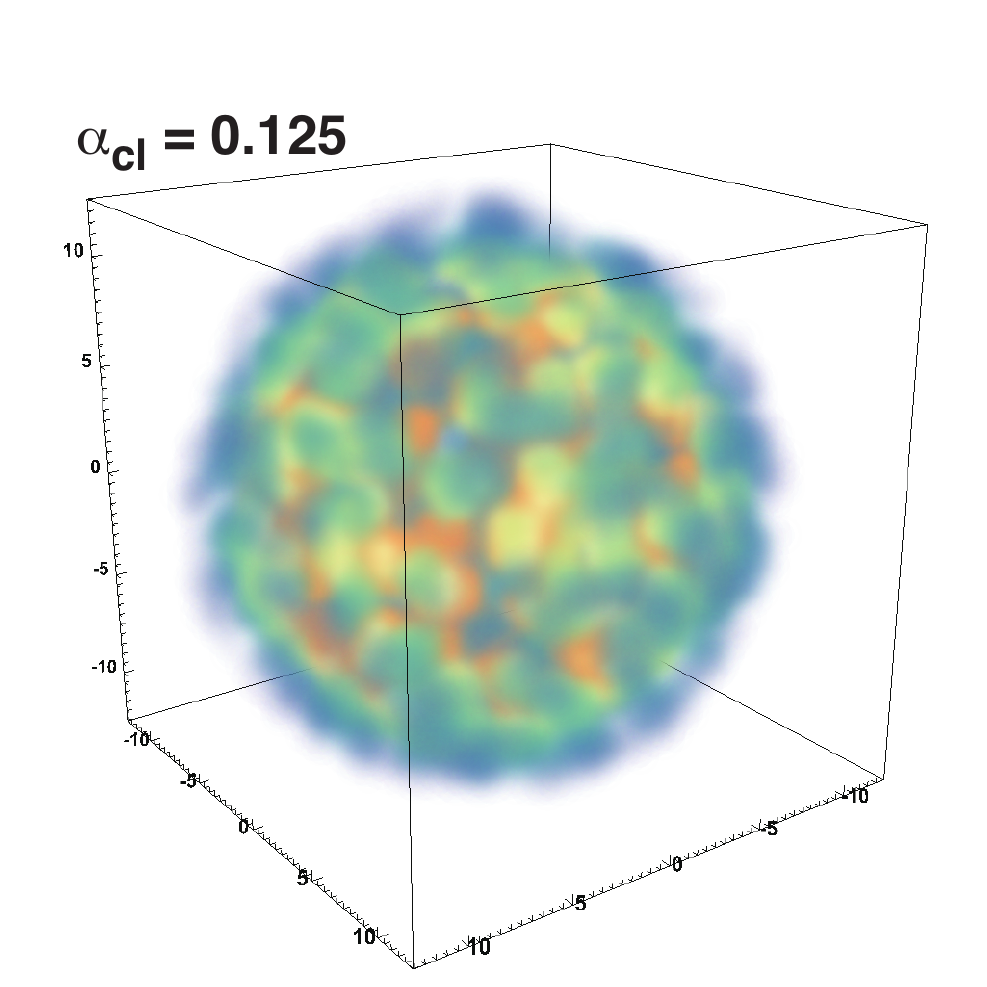} & 
\includegraphics[scale=0.85]{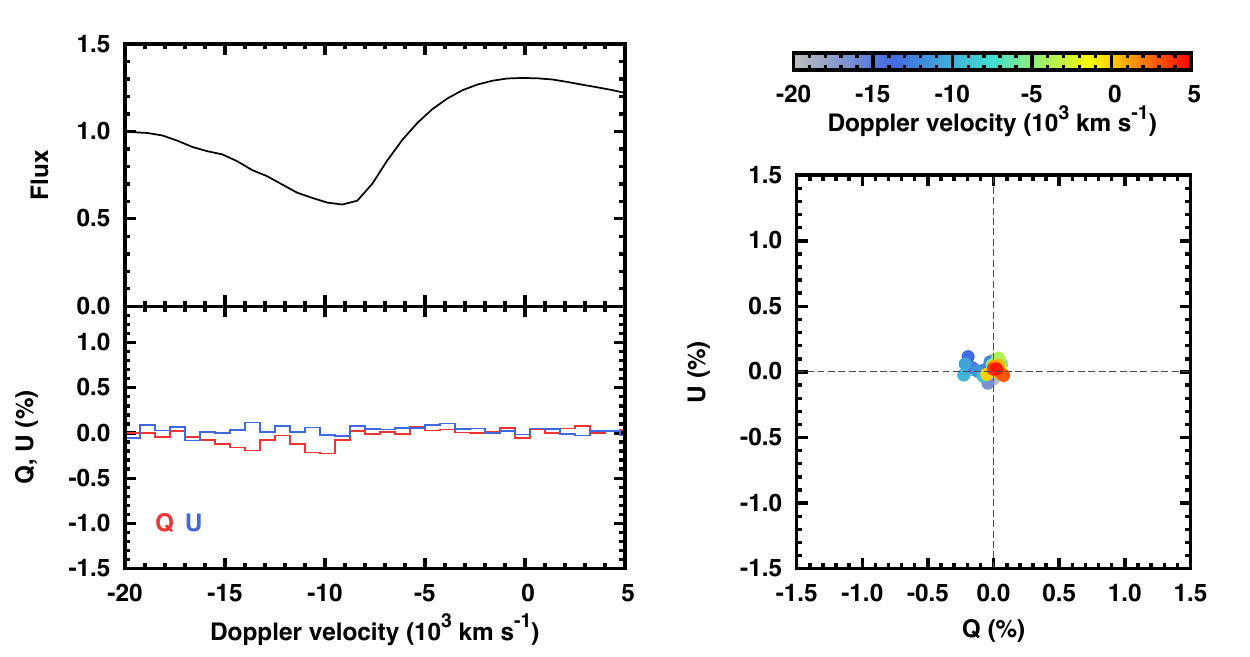}  
\end{tabular}
\caption{
Same with Figure \ref{fig:2D} but for the 3D models.
In the models, the radius of the clumps are 
set to be $\acl = $ 0.5 (top, 3D-a0.5-f0.3), 0.25 (middle, 3D-a0.25-f0.3), 
and 0.125 (bottom, 3D-a0.125-f0.3),
by keeping the covering factor to be $\fcl = 0.3$.
A line of sight for the polarization spectrum is randomly selected.
\label{fig:3D}}
\end{center}
\end{figure*}

\section{Results: 3D Models}
\label{sec:3D}

Next we study polarization properties of 3D models.
Motivated by 3D simulations of neutrino-driven explosions,
where various sizes of complex structure appear, 
we set up 3D models by randomly placing different numbers of spherical clumps
with different sizes.
Here we introduce two parameters to depict the model:
the size parameter of the clumps $\acl$,
\ie the radius of the clump normalized by the photospheric radius
($\acl = \vcl/\vph$) 
and the photospheric covering factor ($\fcl$).
Since the optical depth near the photosphere is the most important 
for line formation, the covering factor is evaluated 
by taking into account the clumps only in a shell between 
$v = v_{\rm ph}$ and $v_{\rm ph} + 2000$ \kms.
Note that, as in the 2D cases, the line optical depth in our models has a spherical component
and it is enhanced by a factor of $f_{\tau} = 10.0$ within the clumps.

The top panels of Figure \ref{fig:3D} show
the polarization properties of the 3D model with
the clump size of $\acl = 0.5$ and 
the covering factor of $\fcl = 0.3$.
In the polarization spectrum, both Stokes $Q$ and $U$ parameters
vary across the lines (middle panel), and 
polarization shows a loop in the $Q-U$ diagram (right panel),
as also found by \citet{hole10}.

The $Q-U$ loop in the 3D clumpy models can be understood as follows.
In the 3D clumpy models, depending on the Doppler velocities, 
different parts of the photospheric disk are hidden by the clumps.
Since the distribution of the clumps does not have a common symmetric axis, 
the position angle of the polarization can change 
depending on the Doppler velocities.
In general, the change in the position angle across the line 
can be arbitrary large,
\ie the polarization in the $Q-U$ diagram can be scattered around.
But for the relatively large size of the clumps as in the case of $\acl = 0.5$,
the same clump keeps contributing to the absorption 
even for different Doppler velocities,
and thus the change in the position angle tends to be smooth 
as a function of Doppler velocities.
Therefore, the polarization tends to show a loop in the $Q-U$ diagram
in the 3D clumpy distribution with relatively large clumps.
Note that the $Q-U$ loop can also be produced by other geometries,
\eg a combination of the ellipsoidal photosphere and
ellipsoidal line scattering shell
whose symmetric axes are misaligned with each other \citep{kasen03},
But even in such a case, it is required that
the axisymmetry of the system is broken.

\begin{figure}
  \begin{center}    
          \includegraphics[scale=0.5]{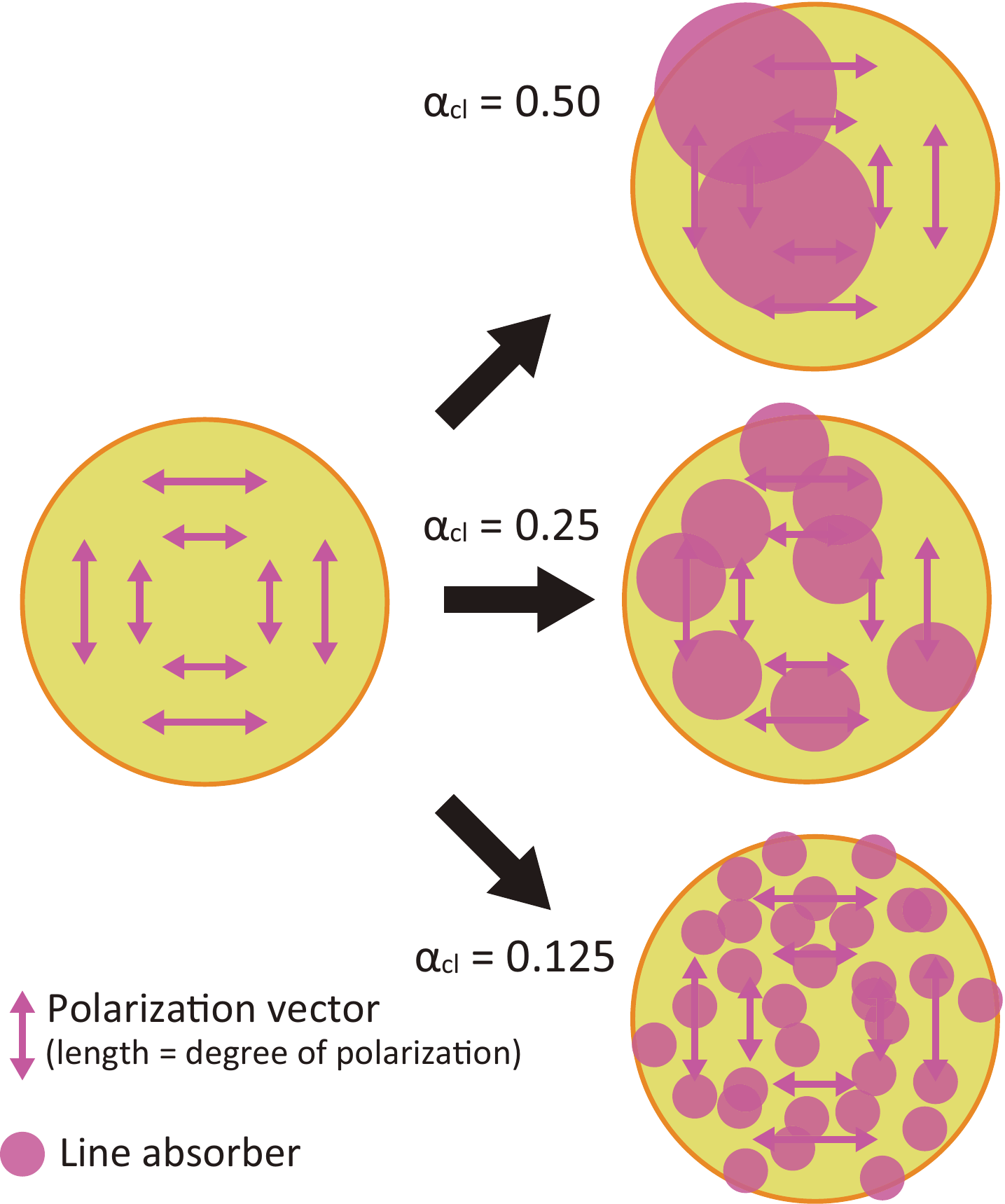}
\caption{
Schematic illustration for SN polarization.
For spherical photosphere (left), polarization vectors are 
cancelled out, and no polarization would be observed in the continuum light.
At the wavelength of absorption lines,
if the distribution of the absorbers (or clumps) is not spherically symmetric,
the cancellation becomes incomplete, and line polarization
would be observed (top).
When the clump is too small (bottom), however, 
polarization vectors tend to be cancelled 
and polarization degree becomes smaller.
\label{fig:schematic}}
\end{center}
\end{figure}

\begin{figure}
  \begin{center}    
  \includegraphics[scale=1.1]{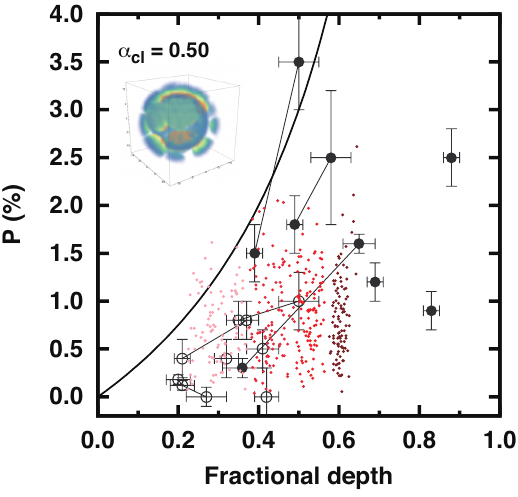}\\
  \includegraphics[scale=1.1]{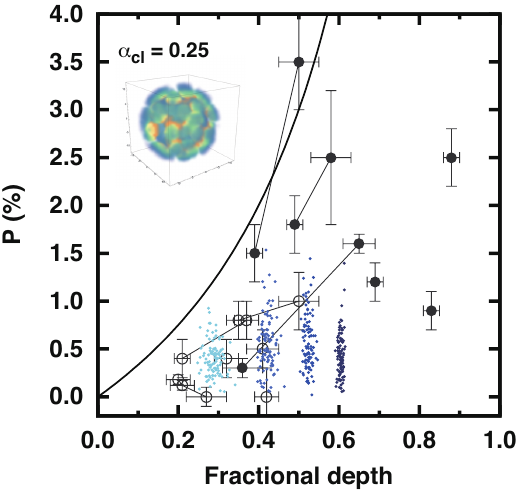}\\
  \includegraphics[scale=1.1]{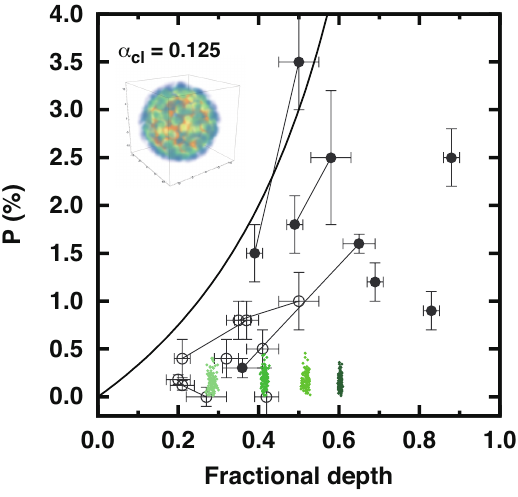}  
\caption{
Polarization degree as a function of the fractional absorption depth.
Small dots in colors show the computed polarization degree for models
with $\acl = 0.5$ (top, red), $\acl = 0.25$ (middle, blue), and $\acl = 0.125$ (bottom, green).
For each panel, four different colors (lighter to darker colors from left to right) 
represent models with four different line strengths at the photosphere
($\tau_{\rm ph}$ = 3.0, 10.0, 30.0, and 100.0, respectively).
For each model, results of 100 lines of sight are shown.
The black points with error bars are observational data 
of the \ion{Ca}{ii} (filled) and \ion{Fe}{ii} (open) lines
for 6 Type Ib and Ic SNe analyzed in \citet{tanaka12pol}:
Type Ib SNe 2005bf \cite{maund0705bf,tanaka0905bf}, 
2008D \cite{maund09}, 2009jf \cite{tanaka12pol}, Type Ic SNe 2002ap 
\cite{kawabata02,leonard0202ap,wang0302ap}, 2007gr \cite{tanaka0807gr},
and 2009mi \cite{tanaka12pol}.
The solid line shows $P_{\rm obs} = 3.0 \% \times [{\rm FD}/({\rm 1-FD})]$
(see Section \ref{sec:methods}).
\label{fig:Pobs_FD}}
\end{center}
\end{figure}

\begin{figure}
  \begin{center}    
          \includegraphics[scale=1.4]{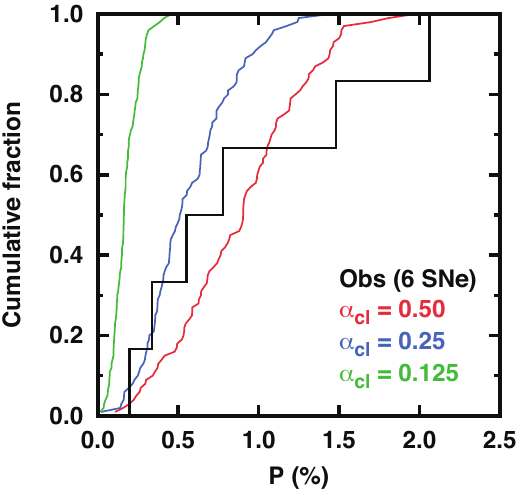}
\caption{
Cumulative distribution of polarization properties of 6 Type Ib and Ic SNe
in Figure \ref{fig:Pobs_FD}.
One characteristic polarization degree is assigned for each object 
by taking the average of the corrected polarization 
($P_{\rm cor}$) of the \ion{Ca}{ii} and \ion{Fe}{ii} lines.
Color lines show the cumulative distribution of polarization degree
for 100 lines of sight.
Three models with
$\acl = 0.5$ (red), $\acl = 0.25$ (blue), and $\acl = 0.125$ (green)
are shown.
In this plot, we use the models with $\tau_{\rm ph} = 30.0$ since
these models approximately give $FD \sim 0.5$ (Figure \ref{fig:Pobs_FD}),
where the corrected polarization is defined.
\label{fig:Pdist_size}}
\end{center}
\end{figure}

\begin{figure*}
\begin{center}
      \includegraphics[scale=1.3]{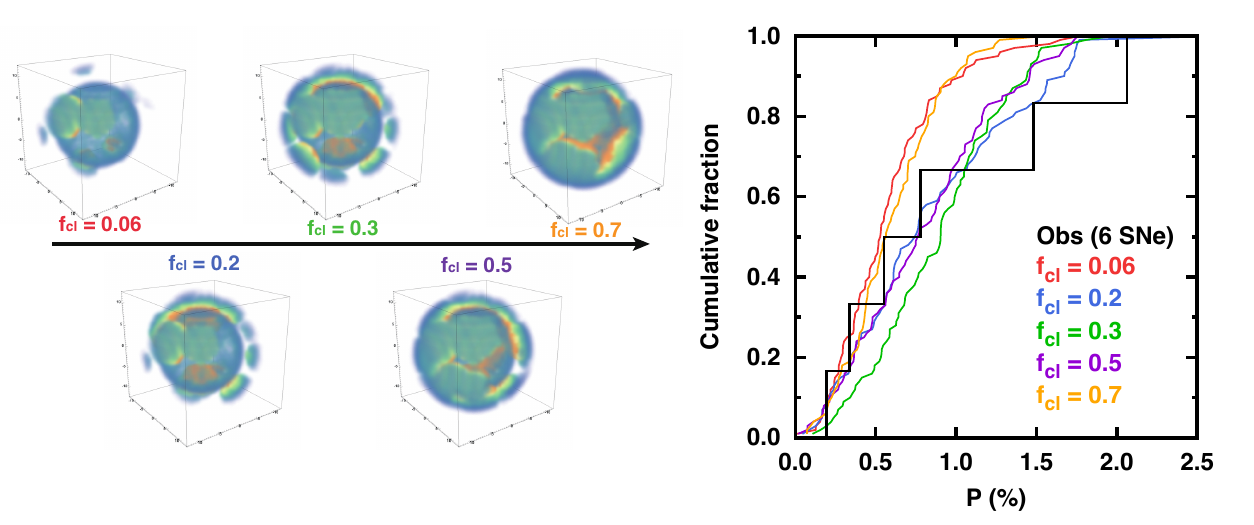} 
\caption{
  {\it Left}:
  Distribution of the optical depth for the models with
  different covering factors.
  {\it Right}: Cumulative distribution of observed polarization
  and the models.
\label{fig:Pdist_fc}}
\end{center}
\end{figure*}

\subsection{Size of the clumps}
\label{sec:3D_size}

The size of the clumps is of interest to 
study the origin of the 3D structure in the SNe.
We show the first attempt to quantify the size of the clumps
by comparing the results of the modelling
and the observed polarization degrees, \ie 
the maximum polarization level at the absorption line.
We calculate the polarization spectra with
different sizes of the clumps by keeping
the covering factor of $\fcl =0.3$ and other parameters to be the same.
The middle and bottom panels in Figure \ref{fig:3D} 
show the results for the 3D models with 
$\acl = 0.25$ and $0.125$, respectively.

As shown in the figures, for a given covering factor,
models with smaller clumps show a lower polarization.
In such models, the photospheric disk is hidden by 
many small clumps, and polarization vectors tend to be 
cancelled out (Figure \ref{fig:schematic}).
This behavior was also pointed out by \citet{hole10} 
in the context of Type Ia SNe.
Since stripped-envelope SNe generally show non-zero line polarization,
the typical size of the clumps should not be too small.

Since the polarization degree depends not only on the geometry 
but also on the absorption strength, it is important to compare 
the models and observations for similar absorption strengths.
Therefore, in Figure \ref{fig:Pobs_FD}, 
we compare models and observations
in the plane of the polarization degree and the fractional absorption depth.
The black points with error bars are observational data 
of the \ion{Ca}{ii} (filled) and \ion{Fe}{ii} (open) lines
for 6 Type Ib and Ic SNe analyzed in \citet{tanaka12pol}.
The small dots show the polarization degree of the models
for 100 lines of sight.
In each panel, we show four sets of the models with the same size and distribution
of the clumps but with the different line optical depth at the photosphere
($\tau_{\rm ph}$ = 3.0, 10.0, 30.0, and 100.0 from left to right).

When the clump is as small as $\acl = 0.125$ 
(bottom panels of Figures \ref{fig:3D} and \ref{fig:Pobs_FD}), 
the polarization degree cannot be $> 0.5 \%$ for any line of sight.
For the larger sizes of the clumps, a higher polarization can be obtained.
When the size of the clumps is $\vcl = 0.25$ (middle panels),
the polarization degrees of these models are still short of 
some of the observed polarization.
When the size of the clumps is relatively large, 
$\acl = 0.5$ (top panels),
the polarization degree can be as high as $> 1 \%$
for the fractional depth of 0.5.

Ideally the polarization properties of the models should be compared
with the statistical distribution of the observed polarization.
Although the number of objects with good data is still small,
Figure \ref{fig:Pdist_size} shows a cumulative distribution
of polarization properties of 6 Type Ib and Ic SNe.
To define one characteristic polarization for each object, 
we take the average of the corrected polarization ($P_{\rm cor}$) 
for the \ion{Ca}{ii} and \ion{Fe}{ii} lines.
Color lines show the cumulative distribution
of the modelled polarization for 100 lines of sight.
We choose models with $\tau_{\rm ph} = 30.0$, 
which approximately give $FD \sim 0.5$ (Figure \ref{fig:Pobs_FD}).

The comparison in the cumulative distribution 
clearly shows that the model with too small clumps ($\acl = 0.125$) 
is not consistent with the observations.
The $p$ value for a Kolmogorov-Smirnov (KS) test is $p_{\rm KS} = 0.0016$.
Since the number of objects is so small,
we cannot distinguish the model with the clump size of
$\acl = 0.25$ ($p_{\rm KS}=0.54$)
and $\acl = 0.5$ ($p_{\rm KS}=0.94$).
Nevertheless, the model with $\acl = 0.25$ is already short to 
explain the polarization level of $>1 \%$, and seems 
to close to the lower limit for the clump size.
Here it is noted that our models adopt
  an enhancement factor of $f_{\rm \tau} = 10.0$.
For a higher enhancement factor, the polarization degree  
is not largely affected because models with $f_{\rm \tau} = 10.0$ already
give an optically thick absorption in the clumps near the photosphere.
On the other hand, for a smaller enhancement factor,
the polarization degree decreases for a given FD.
In such cases, even larger clumps is required
to reproduce a high polarization degree.
Therefore, we conclude that 
a typical size of the 3D clumps should be $\gsim 25\%$ of the photospheric
radius to reproduce observed polarization degrees.

\subsection{Covering factor of the clumps}
\label{sec:3D_cf}

To obtain possible constraints on the number or the covering factor
of the clumps in the ejecta,
we vary the covering factors of clumps keeping their size to be $\acl = 0.5$.
Figure \ref{fig:Pdist_fc} shows the model input (left)
and cumulative distributions of the resultant polarization (right).
For the models, we choose the line strength at the photosphere 
to have $FD \sim 0.5$, \ie
$\tau_{\rm ph}$ = 30 for the models with $\fcl = 0.06$, 0.2, and 0.3.
$\tau_{\rm ph}$ = 10 for the models with $\fcl = 0.4$ and 0.7.
The observed distribution is the same as in Figure \ref{fig:Pdist_size}.

For the model with a smaller covering factor ($\fcl = 0.06$),
the probability to have a high polarization is also low.
Then, by increasing the covering factor of the clumps,
a higher polarization can be more frequently observed ($\fcl = 0.2-0.5$).
However, if the covering factor of the clumps is too large ($\fcl = 0.7$),
the distribution of resultant polarization shifts toward a lower value again
since the system restores the symmetry again.

Since the observational samples are small, 
it is difficult to draw a firm conclusion on the covering factor
of the clumps.
However, the models with $\fcl = 0.06$ and $\fcl = 0.7$ 
are already at the edge of the distribution.
By taking into account the fact
  that models with $f_{\rm \tau} = 10.0$ tend to give an upper limit of
  the polarization level (see Section \ref{sec:3D_size}),
it seems that current data do not support
models with too small covering factors ($\fcl \lsim 0.05$) 
and too large covering factors ($\fcl \gsim 0.8$).

\section{Discussion}
\label{sec:discussion}

We have modelled line polarization of stripped-envelope core-collapse SNe.
The results of modelling are summarized as follows.
(1) The observed $Q-U$ loop cannot
be explained by the 2D axisymmetric models,
but can be explained by the 3D clumpy models.
(2) By comparing the results of the 3D clumpy models with
the observed degrees of line polarization,
it is found that a typical size of the clumps is relatively large,
\ie $\gsim 25\%$ of the radius, and 
a covering factor of the clumps in the ejecta is not
too small and not too large ($5-80\%$).
    
It is intriguing that such a large-scale clumpy structure
is also seen in the element distribution of Cassiopeia A
\citep[\eg][]{hwang04,isensee10,delaney10,grefenstette14,milisavljevic15},
which is a supernova remnant produced by a Type IIb SN \citep{krause08}.
The similarity suggests that the element distribution
as seen in Cassiopeia A may also be able to reproduce
the polarization properties observed in early phase of SNe.
Here we discuss possible origins for the clumpy structure
suggested by observations and modelling.

One scenario is the Rayleigh-Taylor (RT) instability,
causing matter mixing in the SN ejecta.
By the RT instability, many clumps are produced 
and metal-rich ejecta inside are delivered toward the outer layers
\citep[\eg][]{hachisu90,fryxell91,herant94b,nagataki98,kifonidis03,joggerst10,ono13}.
However, the RT instability alone usually produces 
small fingers in many directions.
This is similar to the case of $\acl = 0.125$ in Figure \ref{fig:schematic}
and not consistent with the observations.

The clumpy structure suggested by observations 
is more in favor of large-scale convection or SASI
developed in the initial stage of the explosion.
When the large-scale convection or SASI takes place, 
the subsequent evolution of the shock becomes asymmetric, 
which produces the large-scale asymmetry in the element distribution 
\citep[\eg][]{kifonidis03,kifonidis06,hammer10,fujimoto11}.
Also, results of long-term simulations show that
the ejecta structure near the shock breakout still keeps 
an imprint of the large-scale asymmetry generated
by neutrino-driven convection and SASI,
with the small-scale structures by the RT instability added on top of it
\citep{wongwathanarat13,wongwathanarat15}.
Note that such long-term simulations for neutrino-driven explosion
also nicely reproduce the geometry of Cassiopeia A \citep{wongwathanarat16}.

It is worth noting that,
although the loop in the $Q-U$ diagram does not support 
a purely axisymmetric element distribution (Figure \ref{fig:2D}),
spectropolarimetric data do not rule out
the presence of an overall bipolar structure
or a dominant axis in the SN ejecta.
As long as some large-scale, non-axisymmetric components exist,
they can produce a large enough polarization level
and the loop in the $Q-U$ diagram.
In fact, analysis of the [\ion{O}{i}] line profiles in the late-phase spectra
suggest a torus-like distribution of oxygen,
which is consistent with a bipolar explosion
\citep[\eg][]{maeda08,modjaz08,tanaka0908Dneb}.
Since polarization at the early phase and nebular line profile
are sensitive to the outer and inner ejecta, respectively,
the combination of early and late phase observations
may indicate that global 2D structure exists
more in the inner ejecta and 3D clumpy structure is added in the outer ejecta.
It is noted that, even by the late phase observations,
presence of clumpy structure has also been suggested by the studies of 
line profiles \citep[\eg][]{spyromilio94,sollerman98,taubenberger09},
ionization states \citep[\eg][]{mazzali07,mazzali0706aj},
and dust \citep[\eg][]{sugerman06,ercolano07,kotak09,wesson15,dwek15,bevan16}.
Thus, the transition from the inner 2D to
the outer 3D structures may be somewhat gradual.
Interestingly, our studies suggest that the shape of the loop in the $Q-U$ diagram can 
be used as a probe of such a combined (2D $+$ 3D) geometry.
As expected from the results of 2D  (Figure \ref{fig:2D}) and 3D models (Figure \ref{fig:3D}),
if the ejecta has an overall 2D geometry + 3D clumpy structure,
it tends to produce an elongated loop in the $Q-U$ diagram.
Although current observational data do not allow us to extract such information,
detailed studies will be possible in future with more observational samples 
with high signal-to-noise ratio.

It is emphasized that our modelling includes many simplifications.
For example, we parameterize the line optical depth
and enhancement factor,
but they must be determined by the combination of element abundance,
temperature, and ionization states.
Thus, our models shown in the left panels of 
Figures \ref{fig:2D} and \ref{fig:3D} are not readily 
connected with the element distribution.
Full radiation transfer modelling using
3D hydrodynamic models 
is required to obtain a closer link between the explosion models
and observations.
Also, the comparison with observed polarization degree is done 
by averaging the polarization degrees of different lines.
Since polarization at different absorption lines reflects the distribution of each element and ion,
the direct comparison for each element is necessary when 
larger observed samples and full transfer calculations are available.

\section{Summary}
\label{sec:summary}

We have performed 3D radiation transfer simulations
for the analysis of line polarization in stripped-envelope SNe.
We demonstrate that a purely axisymmetric, 2D structure always produces
a straight line in the Stokes $Q-U$ diagram, and cannot
explain the commonly observed loop in the $Q-U$ diagram.
On the contrary, 3D clumpy structures naturally reproduce the loop.
Comparison of the results of the modelling and
the observed polarization degrees enables 
to constrain a typical size of the clumps from polarization data
for the first time.
To reproduce the distribution of the observed
polarization degrees (0.5-2.0 \%),
a typical size of the clump should be relatively large,
\ie $> 25\%$ of the photospheric radius 
(or the radius where the clump is located).
The covering factor of the clump in the ejecta is only weakly constrained
\ie to $5-80$ \%.
Such a large-scale clumpy structure inferred by polarization
is similar to that seen in the SN remnant Cassiopeia A.

The large-scale clumpy structure is unlikely to be produced 
only by the RT instability as it tends to produce
small fingers in many directions.
Instead, the presence of the large-scale clumpy structure 
in the ejecta suggests that 
large-scale convection or SASI takes place at the onset of the explosion.
Polarization properties do not necessarily exclude 
the presence of a dominant axis in the SN ejecta
since non-axisymmetric structure on top of the 2D axisymmetric structure
can also reproduce the loop in the $Q-U$ diagram.
In fact, the analysis of the nebular spectra supports
a bipolar geometry in the innermost layer.
These observational constraints suggest that
SN ejecta may have an overall 2D bipolar
structure inside and 3D clumpy structure outside.
We speculate that such a hybrid structure could be produced by SASI.
In order to obtain further constraints on the explosion mechanism,
polarization modelling using realistic SN models will be worthwhile
as more and more long-term realistic simulations
from core collapse to the shock breakout are becoming available.

\acknowledgments
We thank Takashi Hattori, Kentaro Aoki, Masanori Iye,
Elena Pian, Toshiyuki Sasaki, and Masayuki Yamanaka
for their contribution to the spectropolarimetric observations
with the Subaru telescope, and the referee for valuable comments.
MT thanks Thomas Janka, Takashi Moriya, and Takaya Nozawa
for fruitful discussion.
Numerical simulations presented in this paper 
were carried out with Cray XC30 at Center for Computational Astrophysics, 
National Astronomical Observatory of Japan.
This research has been partly supported by the Grant-in-Aid
for Scientific Research from JSPS
(24740117, 26800100, 15H02075) and MEXT (25103515, 15H00788), and
by World Premier International Research Center Initiative (WPI Initiative),
MEXT, Japan.


\appendix

\section{Three-Dimensional Radiation Transfer Code}

We have developed a new 3D radiation transfer code
to compute polarization spectrum of one line
from arbitrary 3D distribution of the line optical depth.
The code uses the Monte Carlo method,
which is a common method to compute 
polarization by scattering processes
\citep[\eg][]{daniel80,whitney92,hillier91,code95,whitney11}.
For the application to SNe, see \citet{hoeflich91,kasen03,kasen06,dessart11pol}.

\subsection{Spatial and Wavelength Grid}
We set up the 3D Cartesian spatial mesh with the
$100 \times 100 \times 100$ meshes.
The velocity is used as a spatial coordinate 
because the SN ejecta are homologously expands ($r \propto v$),
The outer velocity of the grid is $v_{\rm max} =$ 25000 \kms, 
and thus, the resolution is $\Delta v =$ 500 \kms.
This spatial resolution gives the wavelength resolution
of $\lambda/ \Delta \lambda = c/ \Delta v = 600$,
which is sufficient to make comparison with observed data.

Since the code computes only one (arbitrary) line, 
the wavelength range used in the computation is very small.
If the rest wavelength of the line is $\lambda_0$, 
we compute the spectrum only at the wavelength 
range between
$\lambda_0 (1-v_{\rm max}/c)$ and $\lambda_0 (1+v_{\rm max}/c)$.
Within this wavelength range, the energy spectrum is 
assumed to be constant ($\lambda F_{\lambda}$ = const).

\subsection{Beginning of the Simulations}
Our code assumes a sharply defined inner boundary,
and solves radiation transfer above the boundary
by tracking the photon packets in the expanding ejecta.
Every photon packet has assigned energy, wavelength, 
and Stokes parameters.
Especially each photon packet in the simulation has 
a constant energy, 
irrespective of the wavelength of the packet.
Because of this treatment, 
any photon is not lost during the simulation,
which results in the accurate energy conservation 
\citep{lucy99,kasen06,kromer09}.

The position of the inner boundary is determined so that
the electron scattering optical depth from the inner boundary
to infinity is $\tau_{\rm in}$.
In the simulations used in the main text of the paper,
we always adopt $\tau_{\rm in} = 3$ (Table \ref{tab:param})
as in \citet{kasen03} and \citet{hole10}.
The radiation from the inner boundary is assumed to be thermalized,
and thus, to be unpolarized;
\begin{equation}
\bm{I} = 
\left( \begin{array}{c}
I\\ Q\\ U
\end{array} \right)
= 
\left( \begin{array}{c}
1\\ 0\\ 0
\end{array} \right).
\end{equation}
The direction of the photon is determined by 
$\mu = \sqrt{z}$ \citep{mazzalilucy93}
(hereafter we use $z$ to denote a random number, $0 < z \leq 1$), 
where $\mu$ is cosine of the angle 
between the radial and photon direction.
The azimuthal angle around the radial direction $\psi$
is uniformly distributed, \ie $\psi = 2 \pi z$.

\subsection{Scattering Events}
The emitted photon packets experience the electron scattering 
and the line scattering, which are treated in a similar 
way to that by \citet{mazzalilucy93}.
For the electron scattering, we assume a power-law density 
structure with the power-law index $n$.
We also have the photospheric velocity ($\vph$) and 
the epoch from the explosion ($\td$) as input parameters.
The photospheric radius ($r_{\rm ph} = \vph \td$) 
is defined to be the radius where the optical depth 
for the electron scattering is unity.
By setting $\vph$ and $\td$, the normalization of 
the electron density is determined.

For the line scattering, we use the Sobolev approximation \citep{castor70},
and assume a power-law optical depth profile with 
the same index $n$.
The parameter for the line scattering is $\tau_{\rm ph}$,
the optical depth at the photosphere.
In addition, we assume enhancement of the optical depth
by a factor of $f_{\tau}$ in some region.
The parameters used in the simulations are summarized in 
Table \ref{tab:param}.

A photon packet propagating in one computational grid
can have 3 possible events;
(1) escaping from the grid, 
(2) the electron scattering,
and (3) the line scattering.
The event that actually occurs is judged by calculating 
the length to the 3 events.
It is simple to compute the length to the next grid $l_{\rm grid}$
for the given position and the direction vector of 
the photon packet.
The direction to the electron scattering event is
computed by the randomly selected event optical depth
$\tau_R = -\ln(z)$.
When the optical depth reaches this value, 
a scattering event occurs.
Thus, the distance to the electron scattering 
$l_{\rm elec}$ can be computed by 
$\tau_R = n_{e}(\bm{r}) \sigma l_{\rm elec}$.
When $l_{\rm elec}$ is shorter than $l_{\rm grid}$, 
the electron scattering occurs if there is no 
contribution of line scattering.

Since the line scattering is treated as a resonance,
the distance to the line scattering event is 
$l_{\rm line} = c\td (\lambda_0 - \lambda')/\lambda_0$,
where $\lambda'$ is the comoving wavelength of the 
photon packet.
If $l_{\rm line}$ is shortest among 3 lengths, 
the line scattering is taken into account.
The line scattering event actually occurs when 
the sum of the line scattering optical depth ($\tau_{\rm line} (\bm{r})$) 
and the electron scattering optical depth in $s_{\rm line}$ 
($\tau_e = n_{e}(\bm{r}) \sigma l_{\rm line}$) exceeds $\tau_R$.
If this sum does not reach $\tau_R$, then 
the electron scattering opacity is evaluated and added again,
and the fate of the packet is the electron scattering
or the escape from the grid.
For illustration of this process, see Figure 1 of \citet{mazzalilucy93}.

When the scattering event occurs, 
the next direction vector of the photon packets is determined.
For the electron scattering, this scattering 
angle depends on the polarization, which
is discussed in the next Section.
For the line scattering, the direction is 
determined by the isotropic probability function 
in the comoving frame.

By the scattering event, the energy and the wavelength 
of the packet are changed.
For the energy, by the energy conservation in the rest frame, 
\begin{equation}
\epsilon_{\rm out} = \epsilon_{\rm in} \frac{1 - \mu_{\rm in}v/c}{1 - \mu_{\rm out}v/c},
\end{equation}
where $\epsilon_{\rm in}$ and $\epsilon_{\rm out}$
are the rest-frame energy of the incoming and outgoing packets, respectively.
And $\mu_{\rm in}$ and $\mu_{\rm out}$ are the cosines of the angles 
between the radial direction and incoming/outgoing
propagating directions, respectively.
Similarly, the change in the wavelength is given by
\begin{equation}
\lambda_{\rm out} = \lambda_{\rm in} \frac{1 - \mu_{\rm out}v/c}{1 - \mu_{\rm in}v/c},
\end{equation} 
where $\lambda_{\rm in}$ and $\lambda_{\rm out}$ are the rest-frame 
wavelength of the incoming and outgoing packet, respectively.

\subsection{Polarization Calculations}

Scattering events change the polarization properties of 
the photon packets.
For the electron scattering, the phase matrix 
can be written as follows \citep{chandrasekhar60};
\begin{equation}
\bm{R}(\Theta) = \frac{3}{4}
\left( \begin{array}{ccc}
\cos^2 \Theta + 1 & \cos^2 \Theta - 1 &  0 \\
\cos^2 \Theta - 1 & \cos^2 \Theta + 1 &  0 \\
0                 &        0          &  2 \cos \Theta
\end{array} \right),
\end{equation}
where $\Theta$ is the scattering angle on 
the plane of the scattering.
This matrix should be operated in the scattering frame.
In general, the rotation matrix for the Stokes parameters
is written as follows \citep{chandrasekhar60};
\begin{equation}
\bm{L}(\phi) =
\left( \begin{array}{ccc}
   1    &   0        &   0        \\
   0    & \cos 2\phi & \sin 2 \phi \\
   0    & -\sin 2\phi & \cos 2 \phi \\
\end{array} \right).
\end{equation}
By using these matrices, the effect 
on the Stokes parameters is given by
\begin{equation}
\bm{I}_{\rm out} = \bm{L}(\pi - i_2) R(\Theta) \bm{L}(-i_1) \bm{I}_{\rm in}.
\label{eq:pol_escat}
\end{equation}
Here $\bm{I}_{\rm in}$ and $\bm{I}_{\rm out}$ is Stokes parameter
in the rest frame before and after the scattering, respectively.
The angles $i_1$ and $i_2$ are the angles on the 
spherical triangle defined as in 
(\citealt{chandrasekhar60}, see Figure 1 of \citealt{code95}).

Equation \ref{eq:pol_escat} means that the the angle-dependence 
of the intensity of the scattered light depends the 
polarization properties of the incident radiation.
From Equation \ref{eq:pol_escat}, 
the probability distribution function ($p.d.f$) 
of the total intensity is
\begin{equation}
p.d.f = \frac{1}{2} (\cos^2 \Theta + 1) + \frac{1}{2} (\cos^2 \Theta -1)
(\cos 2i_1 Q_{\rm in}/I_{\rm in} - \sin2i_1 U_{\rm in}/I_{\rm in}).
\end{equation}
By using this function with the rejection method as outlined 
in \citet{code95},
we determine the scattering angle of the electron scattering.

We assume that the line scattering
works as a depolarizer, \ie the emission turns into 
unpolarized state by the line scattering
as assumed in previous studies 
\citep[see][]{hoeflich96,kasen06,hole10}.

\subsection{Test Calculations}

For the computation of polarization for the electron scattering,
the code was tested with the analytic formulae
by \citet{brown77} for optically thin cases,
and with numerical results by \citet{code95} for optically thick cases. 
For both cases, we got an excellent agreement.
For the application to a SN, expanding ejecta
with the steep density slope, we checked
our results with those by \citet{kasen03}.
We confirmed that our code gives the consistent results 
on the radial profile of polarization for several power-law indexes ($n$) and
the inner boundaries ($\tau_{\rm in}$).

\end{document}